\documentclass[pre,aps,twocolumn,showpacs,amsmath,amssymb,amsfonts]{revtex4}
\usepackage{epsfig}
\usepackage{amscd}
\usepackage{color}
\begin{document}
\title{Three different routes from the directed Ising to the directed
percolation class}
\author{Su-Chan Park}
\affiliation{Institut f\"ur Theoretische Physik, Universit\"at zu K\"oln, Z\"ulpicher Strasse 77, 50937 K\"oln, Germany}
\author{Hyunggyu Park}
\affiliation{School of Physics, Korea Institute for Advanced Study, Seoul
130-722, Korea}
\date{\today}
\begin{abstract}
Scaling nature of absorbing critical phenomena is well understood for the
directed percolation (DP) and the directed Ising (DI) systems. However, a full
analysis of the crossover behavior is still lacking, which is of
our interest in this study. There are three different routes from the DI
to the DP classes by introducing a symmetry breaking
field (SB), breaking a modulo 2 conservation (CB), or making channels
connecting two equivalent absorbing states (CC). Each route can be
characterized by a crossover exponent, which is found numerically as
$\phi=2.1\pm 0.1$ (SB), $4.6\pm 0.2$ (CB), and $2.9\pm 0.1$ (CC), respectively.
The difference between the SB and CB crossover can be understood easily in
the domain wall language, while the CC crossover involves an additional
critical singularity in the auxiliary field density with the memory effect to identify itself independent.
\end{abstract}
\maketitle
\section{\label{Section:Introduction}Introduction}
Critical behavior of absorbing phase transitions can be categorized
by universality classes \cite{H00O04}. The most well-known example is the
directed percolation (DP) class which may be (loosely) defined by a classification scheme 
known as the ``DP conjecture'' \cite{J81,G82,GLB89}. The conjecture states that models 
exhibiting a continuous transition into a {\em single} absorbing state should belong to 
the DP class unless symmetry or conservation is involved. 
The DP conjecture is not complete in two folds.
First, some models with infinitely many absorbing states such as the pair contact 
process~\cite{J93} also belong to the DP class, at least in its stationary property, unless 
there is no clear-cut symmetry among absorbing states \cite{HP99,MM99,PP01}. 
Second, the conjecture itself does not resolve the controversy
related to the pair contact process with diffusion (PCPD)
\cite{HH04,KC03,NP04,PP05,H06,PP06}.

However, this conjecture is useful enough to trigger a search for a non-DP
universality class. One can find several universality classes in Ref.~\cite{H00O04}
which do not meet the constraint of the DP conjecture. Among them, this paper interests 
in the system with two equivalent absorbing states (or in general two equivalent groups 
of many absorbing states) \cite{GKvdT84,TT92,KP94,MO94,J94}.
The Ising-like $Z_2$ symmetry between two absorbing states which is sometimes manifest
as a modulo 2 conservation of domain walls in one dimension
renders the system to exhibit a non-DP critical behavior.
This universality class has several names such as
the parity conserving \cite{GKvdT84,TT92}, the directed Ising (DI) \cite{KP94,HKPP98,NPdN99} 
which will be used to refer to this class in this paper,  the generalized voter
class \cite{HCDM05} and so on. In higher dimensions, all these classes become distinct.

The Ising symmetry (or the modulo 2 conservation) {\it per se} is not enough
to  force the DI critical behavior. It requires an infinite dynamic barrier between two 
equivalent (group of) absorbing states \cite{HKPP98,HP99}, which is analogous to the free 
energy barrier between two ordered states in the equilibrium Ising model. A state near one 
absorbing state cannot evolve into the other absorbing state within a finite number of 
successive local changes. In other words, a domain wall (frustration) in a configuration 
generated by pasting two absorbing states cannot disappear by itself, so the system never 
becomes absorbing with a single frustration \cite{HP99,KC03}. 

As an example without this additional feature, consider a one-dimensional system of diffusing
particles with pair annihilation ($2A \rightarrow 0$) and branching of two particles by a 
triplet ($3A \rightarrow 5A$).  We may map this particle model onto the ferromagnetic Ising 
spin model by interpreting a particle as a domain wall of the Ising system as follows:
\begin{equation}
\begin{aligned}
A\emptyset  \leftrightarrow \emptyset A \quad &\longmapsto \quad
\uparrow \downarrow \downarrow \leftrightarrow \uparrow \uparrow \downarrow,\\
AA\rightarrow \emptyset \emptyset  \quad&\longmapsto  \quad
\uparrow \downarrow \uparrow \rightarrow \uparrow \uparrow \uparrow,
\\
\emptyset A A A \emptyset \rightarrow AAAAA  \quad&\longmapsto \quad
\downarrow \downarrow \uparrow\downarrow\uparrow \uparrow
\rightarrow \downarrow \uparrow \downarrow\uparrow \downarrow \uparrow,
\end{aligned}
\label{Eq:mapping}
\end{equation}
where $A$ ($\emptyset$) stands for a particle (vacancy) and $\uparrow$ ($\downarrow$) 
represents a up (down) spin. Clearly the number of domain walls is conserved modulo 2 and 
there are two absorbing states in the spin system (all spins are up or down) which are 
equivalent. However, the states with only one (diffusing) particle are also absorbing in 
the particle model. In the spin language, these extra absorbing states can be obtained by 
connecting two equivalent absorbing states  (for example, $\ldots \uparrow \uparrow 
\uparrow \downarrow \downarrow \downarrow \ldots$), which violate the infinite dynamic 
barrier requirement for the DI universality class mentioned above~\cite{HKPP98,HP99}.
It is numerically shown that this system shares the critical
behavior with the DP not the DI \cite{KC03,PP08}, while a  similar system with 
$2A \rightarrow 0$ and $2A \rightarrow 4A$ belongs to the PCPD class \cite{PHK01}. 
It is claimed that a long-term memory plays a crucial role in differentiating 
these two universality classes \cite{NP04}.

According to the above criterion for the DI class, one can conceive three possible routes 
along which the DI critical behavior crosses over to the DP in one dimension.
Introducing a symmetry breaking field (SB) \cite{PP95,HKPP98,BB96,H97,KHP99}, allowing 
dynamics which breaks the modulo 2 conservation (CB) \cite{KP95,CT96}, or making channels 
connecting two equivalent absorbing states (CC) \cite{HP99,PP01} drives the system into the 
DP class. But these three routes bring about different characteristics in the absorbing 
states. The SB route maintains the absorbing states but breaks the probabilistic balance 
between them. The CB route (for example, by adding $A \rightarrow 0$ or $2A$) invalidates 
the mapping between the spin and the particle model, and ends up with only one absorbing 
state (vacuum). The CC route is accompanied by additional (infinitely many) absorbing 
states  connecting two equivalent absorbing states with both the symmetry and 
the mapping sustained. 

To quantify the difference of these routes, we measure the crossover exponents governing 
the crossover behavior. In Sec.~\ref{Section:Models}, one-dimensional models are introduced 
belonging to the DI class. In Sec.~\ref{Section:Three}, numerical results for the 
crossover exponents along three different routes are reported. We discuss and summarize 
our results in Sec.~\ref{Section:Discussion}.

\section{\label{Section:Models}DI Models}
We choose two DI models for studying crossover from the DI class to the
DP class: branching annihilating walks with two offspring
(BAW2)~\cite{TT92,J94}
and the interacting monomer model (IM) which is a simplified version of
the interacting monomer-monomer model~\cite{PP01}.
We study the macroscopic absorbing phase transitions~\cite{PP08}
inherent in the model. The suitable order parameters will be defined
when we explain each model in detail. For convenience, the order
parameter for different models will be denoted by $\rho$ throughout
the paper. We only study one dimensional lattice systems with periodic boundary conditions.

For the simulation of the BAW2, we take the dynamic branching rule of Ref.~\cite{KP95}.
The simulation algorithm is detailed as follows:
At first, we choose a particle residing at a lattice point randomly
among, say, $N_t$ particles present at time $t$. 
It may hop to a randomly chosen nearest neighbor site with probability $p$.
If the target site is already occupied, two particles are removed
immediately.  With probability $(1-p)$, this selected particle branches two particles along 
\begin{figure}[t]
\includegraphics[width=0.45\textwidth]{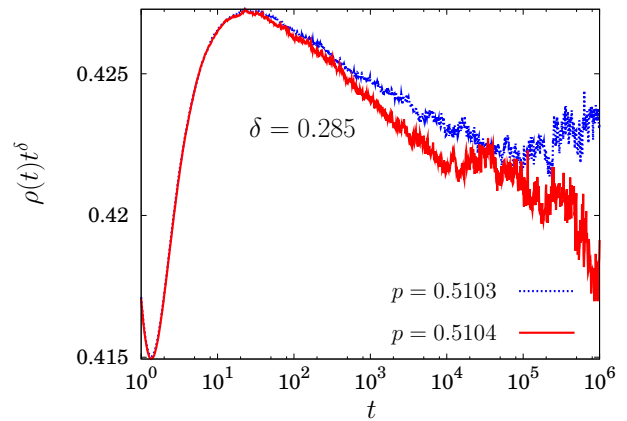}
\caption{\label{Fig:BAW} (Color online) Semilogarithmic plot of $\rho(t) t^{\delta}$ vs $t$
near criticality for the BAW2. The critical point is found to be
$p_c = 0.510\;35(5)$ with the figure in the parentheses as uncertainty
of the last digit.}
\end{figure}
one of two directions.  If an offspring is created on a site already occupied,
both particles annihilate immediately. 
After the above steps, time increases by $1/N_t$.
Since the absorbing state is vacuum, 
the order parameter $\rho$ is the particle density.  At criticality, the density is 
expected to decay as $\rho(t) \sim t^{-\delta}$ with the DI critical exponent
$\delta \simeq 0.285$ \cite{J94}.
Figure \ref{Fig:BAW} locates the critical point at $p_c = 0.510\;35(5)$
which is more accurate than that obtained in Ref.~\cite{KP95}.

The IM is a lattice model 
with two possible states at each lattice point; either vacant
or occupied by a particle. 
The dynamics begins with the adsorption attempt at a randomly
selected vacant site. When both of nearest neighbor (nn) sites are vacant,
a monomer is adsorbed with rate $1$.
In case only one of the nn sites is occupied,
the monomer is adsorbed on the substrate with rate $\lambda$, then
reacts and leaves the substrate with the neighboring particle ($A+A\rightarrow \emptyset$) instantaneously.
When both nn sites are already occupied, the adsorption attempt is rejected. 
The dynamics on the substrate can be summarized as
\begin{subequations}
\label{Eq:ALR}
\begin{eqnarray}
VVV \stackrel{1}{\longrightarrow} VAV,\label{Eq:ALRa}\\
\left . \begin{matrix} AVV\\VVA \end{matrix}
\right \} \stackrel{\lambda}{\longrightarrow} VVV \label{Eq:ALRb},
\end{eqnarray}
\end{subequations}
where $A$ ($V$) means the monomer occupied (vacant) site.

The absorbing states of the IM are characterized by the
``antiferromagnetically ordered'' state
($\ldots AVAVAV \ldots$) and have the $Z_2$ symmetry
(poisoning even or odd sites by monomers). Starting with initial configurations without any $AA$ pair, the proper order parameter $\rho$ is the density of $VV$ pairs.

\begin{figure}[t]
\includegraphics[width=0.45\textwidth]{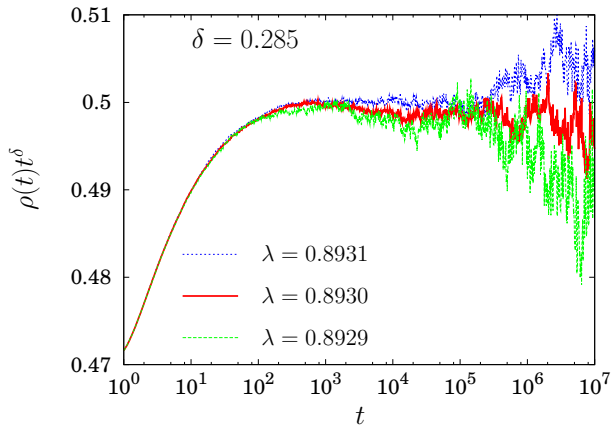}
\caption{\label{Fig:IMc} (Color online) Semilogarithmic plot of $\rho(t) t^{\delta}$ vs $t$
near criticality for the IM. The critical point is found to be
$\lambda_c = 0.8930(1)$.}
\end{figure}

The mapping of the dynamics of the IM to that of
the contact process with the modulo 2 conservation [CP(2)] \cite{IT98}
is possible, though there are some subtleties
which make these two models a little bit different.
If we assign a domain wall excitation, say $X$, in the middle 
of the $VV$ pair, the domain wall dynamics  is exactly the same
as the CP(2) rules.
A pair annihilation in the CP(2) corresponds to an adsorption process
of a monomer as in Eq.~\eqref{Eq:ALRa} and
a pair branching in the CP(2) to an annihilation process Eq.~\eqref{Eq:ALRb}.
Note that the reaction \eqref{Eq:ALRb} always generates two domain walls (two $VV$ pairs), 
because an $AA$ pair is not allowed in the IM.  However, there is no one-to-one 
correspondence between configurations of the CP(2) and the IM, in general. For 
example, a domain wall configuration $X\emptyset X$ is not possible in the IM
by definition, but possible in the CP(2) in principle.  Nevertheless, 
if we start with the fully occupied domain wall state, 
such a configuration cannot be generated by dynamics and thus the mapping becomes exact.
In this case, the parameter $q$ in Ref.~\cite{IT98} for the CP(2)
is related to $1/(1+2\lambda)$ of the IM with suitable time rescaling.
This mapping guarantees that the IM should belong to the DI class
which is confirmed by Fig.~\ref{Fig:IMc}.
We found the critical point of the IM to be $\lambda_c = 0.8930(1)$
and correspondingly the more accurate critical point of the CP(2) as $q_c = 0.358\;94(3)$
than that in Ref.~\cite{IT98}.

\section{\label{Section:Three} Three routes to the DP class}
To study the crossover from the DI class to the DP class, 
we introduce dynamics with rate $w$ which breaks either the symmetry or 
the mod-2 conservation. The crossover scaling ansatz of the order parameter
at criticality is \cite{DL9,PP06}
\begin{equation}
\rho(t;\Delta,w) = t^{-\delta} F(\Delta^{\nu_\|} t, w^{\mu_\|} t),
\label{Eq:sfc}
\end{equation}
where $\Delta$ is the deviation of a tuning parameter from
the DI critical point. The meaning of $\Delta$ and $w$ will be made clearer
in the context of each crossover model.
When $w=0$, the scaling function in Eq.~\eqref{Eq:sfc} should describe the
DI critical behavior. Hence, $\delta$ and $\nu_\|$ are the critical exponents of the DI 
classes; $\delta\simeq 0.285$ and $\nu_\|\simeq 3.2$, respectively \cite{J94}.
The crossover exponent is defined as $\phi = \nu_\| / \mu_\|$
which can be measured from observing how the phase boundary behaves near
the DI critical point
\begin{equation}
1/\phi = \lim_{w\rightarrow 0^+} \frac{\ln{\Delta_c(w)}}{\ln w},
\label{Eq:phi}
\end{equation}
where $\Delta_c(w)$ is the distance from the DI critical point at $w=0$ to the
DP critical point at nonzero $w$.

First, consider the conservation-breaking (CB) route by introducing the mod-2 conservation
breaking dynamics to the BAW2. When a particle tries
to branch offspring, the number of offspring is 2 (1) with
the probability $1-w$ ($w$).
This crossover model with finite $w$ will be referred to as
the BAWCB.

Second, the symmetry-breaking (SB) route is constructed by introducing the symmetry 
breaking dynamics to the IM in such a way
that the adsorption rate Eq.~\eqref{Eq:ALRa} at even-indexed sites
is different from that of odd-indexed sites.
To be specific, we modify adsorption dynamics without affecting the
desorption events as
\begin{equation}
\begin{CD}
VV_\text{o}V @>{1-w}>> VA_\text{o}V,\quad
\end{CD}
\begin{CD}
VV_\text{e}V @>{1}>> VA_\text{e}V,\label{Eq:ALR_SB}
\end{CD}
\tag{\ref{Eq:ALR}a$'$}
\end{equation}
where the subscripts o and e refer to an odd- and even-indexed sites,
respectively.
When $0<w<1$, this system prefers the absorbing state of poisoning monomers
at even-indexed sites, which breaks the $Z_2$ symmetry of the IM.
We will refer to the model with dynamics Eqs.~\eqref{Eq:ALR_SB} and
\eqref{Eq:ALRb} as the IMSB.

\begin{table}[b]
\caption{\label{Table:PB} Critical point values $\lambda_c(w)$ of the IMSB and IMCC  
and $p_{c} (w)$ of the BAWCB
for various $w$'s. The numbers in the
parentheses indicate the uncertainty of the last digits.}
\begin{ruledtabular}
\begin{tabular}{rlll}
$w$&$\lambda_{c}(w;\text{IMSB})$&$\lambda_c (w;\text{IMCC})$&$p_{c}(w;\text{BAWCB})$\\
\hline
0  &0.8930(1) &0.8930(1)&0.510~35(5)\\
$10^{-5}$&&0.9119(1)&0.4931(1)\\
$2\times 10^{-5}$&&&0.490~30(5)\\
$5\times 10^{-5}$&&0.92545(5)&0.4858(1)\\
$10^{-4}$&&0.9340(1)&0.481~65(5)\\
$2\times 10^{-4}$&&0.944~95(5) &0.4767(1)\\
$3\times 10^{-4}$&&&0.473~45(5)\\
$5\times 10^{-4}$&0.9125(2)&0.965~35(5)&0.468~90(5)\\
$10^{-3}$&0.9199(1)&0.9842(1)&0.461~75(5)\\
$2\times 10^{-3}$&0.9303(2)&&\\
$4\times 10^{-3}$&0.944~75(5)&&
\end{tabular}
\end{ruledtabular}
\end{table}
Finally, the channel-connecting (CC) route  can be constructed by
connecting two equivalent symmetric absorbing states through
infinitely many absorbing configurations~\cite{HP99}.
In the context of the IM model, this route can be studied 
by allowing an adsorption without a pair reaction probabilistically:
\begin{equation}
\label{Eq:ALR_mod2}
AVV \stackrel{w}{\longrightarrow} AAV,\quad
VVA \stackrel{w}{\longrightarrow} VAA.
\tag{\ref{Eq:ALR}c}
\end{equation}
The model with dynamics of Eqs.~\eqref{Eq:ALRa}, \eqref{Eq:ALRb}, and
\eqref{Eq:ALR_mod2} will be called as the IMCC.
As an adsorption attempt at a vacant site between two monomers ($AVA$) is still rejected,  
any configuration without a $VV$ pair is absorbing and the order parameter is still the 
$VV$ pair density.  In contrast to the IM, the IMCC has infinitely many absorbing states 
characterized by the $AA$ pair density (auxiliary field density). By pasting the two 
antiferromagnetically ordered absorbing states, we find either $\ldots VAVAAVAV \ldots$ 
which is already absorbing, or $\ldots VAVAVVAVAV \ldots$ which can evolve back to an 
absorbing state by adsorbing a monomer at either site of the $VV$ pair. Hence there is no 
infinite dynamic barrier between absorbing states and the system leaves the DI class into 
the DP class.

Now we are equipped with three different models which can show the crossover behavior from 
the DI class  to the DP class. As explained above, the crossover exponent $\phi$ can be 
deduced from the DP phase boundary near the DI critical point; see Eq.~\eqref{Eq:phi}.
For each model, the critical points at finite $w$ are located numerically using
the DP values of the critical exponents (see Table~\ref{Table:PB}).
The critical points for the IMSB and IMCC are denoted
by $\lambda_c(w)$ and for the BAWCB by $p_c(w)$, respectively.
To measure the crossover exponent, we define  $\Delta_c(w)\equiv | \lambda_c(w) - 
\lambda_c(0) |/\lambda_c(0)$ for the IMSB and the IMCC, and similarly 
$\Delta_c(w) \equiv | p_c(w) - p_c(0) |/p_c(0)$ for the BAWCB.

\begin{figure}[t]
\includegraphics[width=0.45\textwidth]{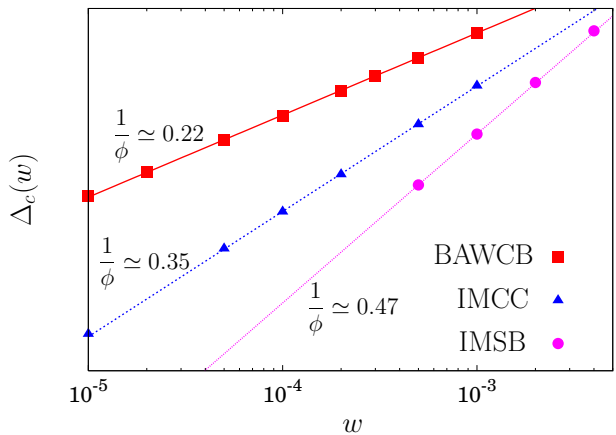}
\caption{\label{Fig:PB} (Color online) Log-log plot of $\Delta_c(w)$ vs $w$
for three different crossover models. For graphical clarity, we shifted $y$
axis with arbitrary scale. Each line whose slope is
$1/\phi$ shows the fitting result for the
corresponding phase boundary.}
\end{figure}

In Fig.~\ref{Fig:PB}, the crossover exponents are estimated for all three different routes. 
Along the SB route, we estimate $\phi=2.1(1)$ which is consistent with 
previous estimates: $\phi=2.1(1)$ \cite{BB96}, $2.24(10)$ \cite{KHP99}, and 
recently $1.9(1)$ \cite{OM08}. Along the CB route, our estimate is $\phi=4.6(2)$ which 
is close to the recent result $\phi= 4.8(2)$ \cite{OM08}. Along the CC route, we 
find $\phi=2.9(1)$, which is clearly different from those for the SB and CB routes.

\section{\label{Section:Discussion} Discussion and Summary}

The crossover behavior from one fixed point to another does not reflect the properties of 
both fixed points, in general. Rather, strictly speaking, it is related to one fixed 
point and its crossover operator which forces the system to crossover to the other 
fixed point \cite{DL9,PP06,PP07}. In many cases, the crossover route is unique between 
two fixed points, so is the crossover operator, which leads to one unique crossover 
exponent between two universality classes. 

However, in the case studied here, we found that there are three different routes for 
the crossover from the DI to the DP class, which yield all three different crossover 
exponents. It indicates that the crossover operator associated with each route should be 
different from each other. The difference between the SB and CB routes has been noticed 
earlier \cite{HKPP98}. Summarizing it in the domain-wall (particle) representation, the 
crossover operator in the CB route is a single particle annihilation or creation operator 
in the action, while that in the SB route is a nonlocal string operator (global product 
of particle number operators) \cite{HKPP98}. Hence it is not surprising to find that the two crossover exponents are different.

\begin{figure}[t]
\includegraphics[width=0.45\textwidth]{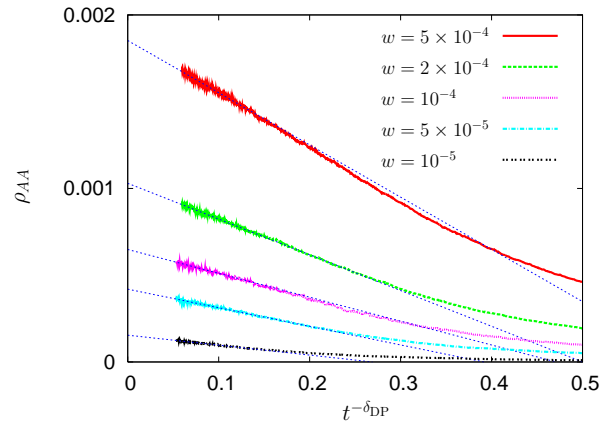}
\caption{\label{Fig:D1} (Color online)
Long-time behavior of the auxiliary field density $\rho_{AA}$ for small $w$'s. 
The exponent $\delta_\text{DP}\simeq 0.160$ is the temporal decay exponent of the 
order parameter in the DP class. Simple extrapolation lines are given by dotted lines.
}
\end{figure}

The crossover behavior along the CC route is tricky. First, note that the $Z_2$ symmetry 
between the odd-indexed and even-indexed sites is not broken in this route, which implies that the CC route is different from the SB one. Second, in 
terms of domain walls, the CC process, Eq.~\eqref{Eq:ALR_mod2}, clearly breaks the mod-2 
conservation in the number of $VV$ pairs (primary field, say, $X$), because
of which one may naively guess the CB-type crossover along the CC route. 
However, unlike the ordinary CB models, the CC process generates 
an infinite number of absorbing configurations with $AA$ pairs (auxiliary field, say, $Y$)
which give the feedback effect to the domain wall dynamics. Consider the dynamics of the IMCC in terms of two fields $X$ and $Y$. The reactions, Eqs.~\eqref{Eq:ALRa} and \eqref{Eq:ALRb}, can be rewritten as $2X\rightarrow\emptyset$ and $X\rightarrow 3X$, while Eq.~\eqref{Eq:ALR_mod2} is equivalent to $X\rightarrow Y$. The presence of $AA$ pairs makes it possible to allow another reaction of $XY\rightarrow 2X$ ($AVVAA\rightarrow AVVVA$) in combination with Eq.~\eqref{Eq:ALRb}. The last two reactions involving the auxiliary field represent the feedback which generates memory effects on the primary field.

\begin{figure}[t]
\includegraphics[width=0.45\textwidth]{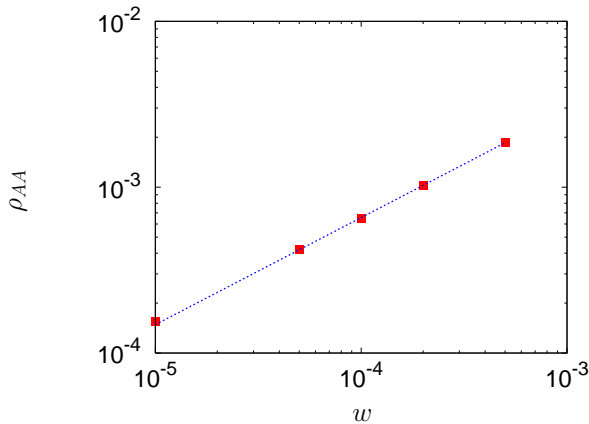}
\caption{\label{Fig:D2} (Color online) 
Log-log plot of  $\rho_{AA}$ in the steady state versus $w$.
The slope of the dotted line is 0.645.
}
\end{figure}

The emergence of the auxiliary field density and the feedback mechanism may be responsible for the new crossover behavior along the CC route. 
To see how the auxiliary field may affect the crossover,
we measure the auxiliary field density (natural density) $\rho_{AA}(w)$ along the DP critical line. 
We expect its long-time temporal behavior as $\rho_{AA}(w,t)=
\rho_{AA}(w) - b t^{-\delta_\text{DP}}$ starting from the $A$ particle vacuum \cite{OMSM98,PP01}, 
where $\delta_\text{DP}$ is the temporal decay exponent of the order parameter in the 
DP class. Figure~\ref{Fig:D1} confirms our expectation and the asymptotic values of 
$\rho_{AA}$ is plotted against $w$ in Fig.~\ref{Fig:D2}. It is 
clear that $\rho_{AA}(0)=0$ at the DI critical point.
Near $w=0$, the power-law singularity is found as $\rho_{AA}\sim w^{\alpha}$ 
with $\alpha= 0.65(2)$. We test the universality of the CC crossover scaling 
as well as the singularity of the auxiliary field density by investigating 
various different models like a CC variant of the interacting 
monomer-dimer model~\cite{PP08b} and a  two-species particle model defined 
as below. It turns out that these models form one crossover universality class with the same CC-type crossover exponent $\phi$ and the auxiliary field density exponent $\alpha$~\cite{PP08b}.

To examine the exclusive role of the feedback (memory) effect, we introduce
a two-species ($X$, $Y$) particle model where the feedback process can be directly controlled. The primary particles $X$ can diffuse, while the secondary particles $Y$ are immobile. Each species particles are hard core particles, but a simultaneous occupation by different species at a site is allowed. Therefore each site can be in one of four possible states such as $\emptyset$, $X$, $Y$, and $XY$. 

The reactions dynamics is  symbolically summarized as $2X\rightarrow \emptyset$, $X\rightarrow 3X$, $X\rightarrow Y$, and $XY\rightarrow Y$, $2X$, or $3X$. 
To be more specific, the evolution rule is given as follows: First, choose one of $X$ particles randomly. With probability $w$, the chosen $X$ spontaneously mutates or annihilates as $X\rightarrow Y$ or $XY\rightarrow Y$. With probability $p-w$, the $X$ hops to one of its neighboring sites. Whenever two $X$ particles attempt to occupy the same site, they annihilate immediately like in the BAW2. With probability $1-p$, the $X$ can branch particles in the neighborhood as $X\rightarrow X+2X$ with the dynamic branching rule~\cite{KP95}. To control the feedback effect, we introduce a parameter $r$ in the branching process when a $Y$ resides along with the $X$ at the same site, such that $XY\rightarrow X+2X$ with the relative rate $(1-r)$ and $XY\rightarrow X+X$ with $r$. 

At $w=0$, there is no spontaneous annihilation (or mutation) process and the model becomes exactly the same as the BAW2 after some transient period needed for removing all $Y$ particles via branching processes. At finite $w$, we expect the DP scaling with a finite density of $Y$ particles; $\rho_Y(w)>0$ because $X$ particles generate $Y$ particles via mutation processes, which cannot be fully eliminated via branching processes. So the CC-type crossover is expected in general. 

However the $r=0$ point is special. There, the presence of $Y$ particles never affects the dynamics of $X$ particles, so there is no feedback mechanism to alter the primary particle density through the secondary particles. The primary field dynamics is basically identical to the BAWCB model introduced in Sec.~\ref{Section:Three} and the CB-type crossover scaling should appear at $r=0$ along  with nonzero $\rho_Y(w)$. At finite $r$, the $X$ particle dynamics is affected by the presence of $Y$ particles, which generates the memory effects on the $X$ particle density. Hence  $r$ can be regarded as the feedback controlling parameter and $w$ as the crossover parameter. 

\begin{figure}[t]
\includegraphics[width=0.48\textwidth]{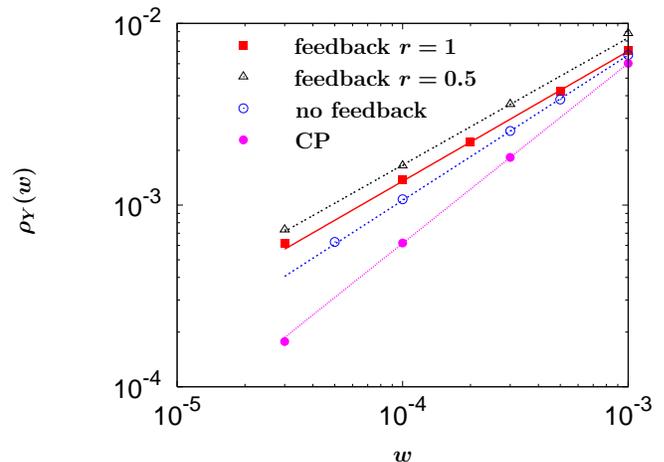}
\caption{\label{Fig:rhoY}(Color online) $\rho_Y(w)$ vs $w$ along the phase boundary for the two-species
model with ($r=1$ and $0.5$) or without ($r=0$) feedback processes as well as the variant of the contact process (see text) in double logarithmic scales. The slopes of the
straight lines are 0.70, 0.71, 0.80, and 1 from above.
}
\end{figure}

We performed numerical simulations for various values of $r$. For nonzero $r$, we found that the secondary particle density behaves as $\rho_Y(w)\sim w^{\alpha}$ with $\alpha= 0.70(3)$ and the crossover exponent $\phi = 3.1(2)$, both of which are consistent with the CC crossover results for the IMCC model within numerical errors. In contrast, at $r=0$
the CB crossover is found as expected along with  $\alpha\simeq 0.80$ different from the CC crossover case; see Fig.~\ref{Fig:rhoY}. This leads to the conclusion that the feedback process (memory) is crucial in establishing the CC-type crossover universality class and affects the singularity characteristic of the auxiliary field density. 

It is interesting to note that the numerical values of $\alpha$ suggest a simple conjecture that $\alpha = 1 - 1/\phi$ for the DI to the DP crossover with an emerging auxiliary field. The mean field analysis leads to $\alpha = 0$ and $\phi = 1$ (see Appendix A) which is also consistent with this conjecture. The nontrivial singularity in $\rho_Y$ is related to the nontrivial crossover from the DI to the DP. Then, it is natural to ask what will happen if there is no nontrivial crossover. To answer this question, we study a two-species version of the contact process (CP) with the secondary particles; $X\rightarrow \emptyset$,  $X\rightarrow 2X$, $X\rightarrow Y$, and $XY\rightarrow Y$, $X$,  or $2X$ with the crossover parameter $w$ controlling the $X\rightarrow Y$ process and the feedback parameter $r$ controlling the $XY\rightarrow X$. This model belongs to the DP class for any value of $w$ and $r$, including $w=0$. Therefore there is no ``true" crossover in this model. We found the trivial linear phase boundary ($\Delta_c(w) \sim w$) as expected and also the trivial behavior of $\rho_Y$ with $\alpha=1$ regardless of the presence of the feedback processes (see Fig.~\ref{Fig:rhoY}). These results are consistent with the mean field results described in the Appendix A.

Up to now, we have been only interested in the emergence of the auxiliary field for a finite crossover parameter $w$ which breaks the mod(2) conservation for the primary particles. One may consider more general cases where the auxiliary field density $\rho_Y$ 
is finite even at the DI critical point ($w=0$). First, we studied the crossover behavior 
to the DP models with a continuous variation of $\rho_Y$ as $w$ increases. We found that $\rho_Y (w)-\rho_Y(0)\sim w^{0.22}$ and the crossover scaling belongs to the CB 
class~\cite{PP08b}. This implies that the vanishing $\rho_Y$ (not just singularity 
near $w=0$) is another important ingredient in the CC crossover. Note that $\rho_Y(w)$ seems to behave in the same way as the phase boundary $\Delta_c(w)\sim w^{1/\phi}$ with $1/\phi\simeq 0.22$ for the CB crossover. 

Second, there may be a discontinuous drop in $\rho_Y$ in the crossover to the DP with
a single absorbing state ($\rho_Y=0$). This case may be compared to the crossover from the 
DP with finite $\rho_Y$ (infinitely many absorbing states) to the DP with $\rho_Y=0$ and 
similarly from the DI with finite $\rho_Y$ to the DI with $\rho_Y=0$~\cite{PP07}. As 
understood in Ref.~\cite{PP07}, this crossover may be characterized by a discontinuous 
jump in the phase boundary at the DI critical point for the excitatory route or 
a continuous phase boundary for the inhibitory route. Along the inhibitory route, the CB 
crossover is observed again (not shown here) and we conclude that the discontinuity in the 
auxiliary field density does not provide any new crossover scaling~\cite{PP08b}. 

To summarize, we studied three routes of the crossover 
from the DI to the DP class;
symmetry breaking (SB), mod-2 conservation breaking (CB), and channel connecting (CC) routes. 
These three routes are characterized by three different crossover exponents. 
The difference between the SB route and the CB route is clear because the symmetry breaking field can be interpreted as the spatial non-local operator affecting the domain wall
dynamics, while the conservation breaking corresponds to the local operator. The CB route and the CC route share some common features, but the feedback memory effect (or temporal non-locality) makes the CC crossover distinct from the CB crossover. 
We also found the universal exponent $\alpha=  0.65(2)$ which describes the singularity of the vanishing auxiliary field in the CC route.
From the numerical study of the feedback controlling model and its mean field theory,
we conjectured that $\alpha = 1 - 1/\phi$.

\acknowledgments
SCP would like to thank for the support of the Korea Institute for Advanced Study (KIAS) 
where this work was initiated and the support by DFG within SFB 680 \textit{Molecular Basis of Evolutionary Innovations}. Most of computation was carried out using KIAS supercomputers.

\appendix
\section{mean field theory}
This appendix provides the mean field theory for the feedback controlling model  introduced in Sec.~\ref{Section:Discussion}.
Here $\rho_X$ and $\rho_Y$ are the densities of $X$ and $Y$ particles, respectively and
$\rho_Z$ is the density of sites where both $X$ and $Y$ particles reside.
Then one may easily write down the mean field equations as
\begin{eqnarray}
\frac{\dot \rho_X}{\rho_X}&=& 2 q - w - 2 ( 1 +   q -  w ) \rho_X
- r q \left (\frac{\rho_Z}{\rho_X} - 2\rho_Z \right ),\label{Eq:X}\\
\dot \rho_Y &=& w \rho_X - (  q + w ) \rho_Z,\label{Eq:Y}\\
\frac{\dot \rho_Z}{\rho_Z} &=& -1 + ( \rho_Y - 2 \rho_Z) 
\left [ (1 +  q - w) \frac{\rho_X}{\rho_Z} - rq \right ],
\label{Eq:Z}
\end{eqnarray}
where $q \equiv 1 -p$. Since $p_c(w=0) = 1$ in the mean field theory, 
$q$ is the same as $\Delta$ in the text.

In the active phase, the steady state density can be calculated by setting $\dot \rho_X = 0$
and so on. The superscript $s$ in the following indicates the steady state density.
From Eq.~\eqref{Eq:Y}, we get in the active phase 
\begin{equation}
\frac{\rho_X^s}{\rho_Z^s} = 1+\frac{q}{w}.
\label{Eq:XZ}
\end{equation}
Hence Eq.~\eqref{Eq:Z} gives the steady state density of $\rho_Y^s$ such that
\begin{eqnarray}
\rho_Y^s &=& 2 \rho_Z^s + \frac{1}{(1+q-w) \rho_X^s/\rho_Z^s -r q }\nonumber\\
&=& 2 \rho_Z^s + \frac{w}{q(1-wr + q) + w(1-w) }.\label{Eq:Y2}
\end{eqnarray} 

In the active phase, the (stable) steady state density $\rho_Y^s$ is determined uniquely,
so we define the natural density at criticality by taking the limiting process from the active side as $\lim_{q \rightarrow q_c^+}\rho_Y^s(q)$.
From Eqs.~\eqref{Eq:X} and \eqref{Eq:XZ}, the critical point is determined
by the equation $- r q + (2 q - w)(1 + q/w)=0$ which reads
\begin{equation}
q_c = \frac{w}{4} \left \{ \sqrt{8 + (1-r)^2} - (1-r) \right \}.
\end{equation}
For any value of $r(\le 1)$, $q_c\sim \Delta_c\sim w$ which gives $\phi=1$. 
Since $\rho_Z^s = 0$ at criticality, we get $\rho_Y^s \sim w/q_c \sim w^0$ from Eq.~\eqref{Eq:Y2}, yielding $\alpha=0$.
In low dimensions, $q_c$ is expected to renormalize such that 
$q_c \sim \Delta_c \sim w^{1/\phi}$, which
leads the conjecture $\alpha = 1 -1/\phi$.

For comparison, we study the two-species version of the contact process with the secondary particles introduced in Sec.~\ref{Section:Discussion}. The mean field equations are
\begin{eqnarray}
\dot \rho_X &=& (2 q - 1 ) \rho_X -  q \rho_X^2 -rq\rho_Z (1 - \rho_X),\\
\dot \rho_Y &=& w \rho_X - ( q + w) \rho_Z,\\
\dot \rho_Z &=& -\rho_Z + q (\rho_Y - \rho_Z) (\rho_X -r \rho_Z),
\end{eqnarray}
where $q=1-p$.
Then one may easily show that the critical line and the critical steady-state density of $Y$ particles are given as for small $w$
\begin{eqnarray}
\Delta_c &\simeq& \frac{r}{2} w, \nonumber\\
\rho_Y^s &=&  \frac{w}{q_c^2 +q_c (1 - r)w}\simeq 4 w ,
\label{Eq:CPMF}
\end{eqnarray}
where $\Delta_c=q_c(w)-1/2$. As expected, 
we find  the trivial phase boundary and also the trivial value of $\alpha=1$, 
which is consistent with the simulation results shown in Fig.~\ref{Fig:rhoY}.
Moreover, the renormalization cannot generate a singular behavior in the denominator
of Eq.~\eqref{Eq:CPMF}, so $\alpha$ is always expected to be 1 in all dimensions.


\begin{thebibliography}{99}

\bibitem{H00O04} H. Hinrichsen, Adv. Phys. {\bf 49}, 815 (2000); G.
\'Odor, Rev. Mod. Phys. {\bf 76}, 663 (2004).
\bibitem{J81} H. K. Janssen, Z. Phys. B: Condens. Matter {\bf 42}, 151 (1981).
\bibitem{G82} P. Grassberger, Z. Phys. B: Condens. Matter {\bf 47}, 365 (1982).
\bibitem{GLB89} G. Grinstein, Z.-W. Lai, D.A. Browne, Phys. Rev. A {\bf 40}, 4820 (1989).
\bibitem{J93} I. Jensen, Phys. Rev. Lett. {\bf 70}, 1465 (1993).
\bibitem{HP99} W. Hwang and H. Park, Phys. Rev. E {\bf 59}, 4683 (1999).
\bibitem{MM99} M. C. Marques and J. F. F. Mendes, Eur. Phys. J. B {\bf 12}, 123 (1999).
\bibitem{PP01} H. S. Park and H. Park, J. Korean Phys. Soc. {\bf 38}, 494 (2001).
\bibitem{HH04} For a review, see M. Henkel and H. Hinrichsen, J. Phys. A {\bf 37}, R117 (2004).
\bibitem{KC03} J. Kockelkoren and H. Chat\'e, Phys. Rev. Lett. {\bf 90}, 125701 (2003).
\bibitem{NP04} J. D. Noh and H. Park, Phys. Rev. E {\bf 69}, 016122 (2004).
\bibitem{PP05} S.-C. Park and H. Park, Phys. Rev. Lett. {\bf 94},065701 (2005); Phys. Rev. E {\bf 71}, 016137 (2005).
\bibitem{H06} H. Hinrichsen, Physica A {\bf 361}, 457 (2006).
\bibitem{PP06} S.-C. Park and H. Park, Phys. Rev. E {\bf 73}, 025105(R) (2006).
\bibitem{GKvdT84} P. Grassberger, F. Krause, and T. von der Twer, J. Phys. A {\bf 17} L105 (1984); P. Grassberger, {\em ibid.} {\bf 22}, L1103 (1989).
\bibitem{TT92} H. Takayasu and A. Yu. Tretyakov, Phys. Rev. Lett. {\bf 68},
3060 (1992).
\bibitem{KP94} M. H. Kim and H. Park, Phys. Rev. Lett. {\bf 73}, 2579 (1994); H. Park, M. H. Kim, and H. Park, Phys. Rev. E {\bf 52}, 5665 (1995).
\bibitem{MO94} N. Menyh\'ard, J. Phys. A {\bf 27}, 6139 (1994); N. Menyh\'ard and G. \'Odor, {\em ibid.} {\bf 29}, 7739 (1996).
\bibitem{J94} I. Jensen, Phys. Rev. E {\bf 50}, 3623 (1994).
\bibitem{HKPP98} W. Hwang, S. Kwon, H. Park, and H. Park, Phys. Rev. E {\bf 57}, 6438 (1998).
\bibitem{NPdN99} J. D. Noh, H. Park, and M. den Nijs, Phys. Rev. E {\bf 59}, 194 (1999).
\bibitem{HCDM05} O. A. Hammal, H. Chat\'{e}, I. Dornic, M. A. Mu{\~n}oz, Phys. Rev. Lett. {\bf 94}, 230601 (2005).
\bibitem{PP08} S.-C. Park and H. Park, Europhys. J. B {\bf xx} xxxx (2008) (DOI:10.1140/epjb/e2008-00022-4).
\bibitem{PHK01} K. Park, H. Hinrichsen, and I.-M. Kim, Phys. Rev. E {\bf 63}, 065103 (2001).
\bibitem{PP95} H. Park and H. Park, Physica A {\bf 221}, 97 (1995).
\bibitem{BB96} K.E. Bassler and D.A. Browne, Phys. Rev. Lett. {\bf 77}, 4094 (1996); Phys. Rev. E {\bf 55}, 5225 (1997).
\bibitem{H97} H. Hinrichsen, Phys. Rev. E {\bf 55}, 219 (1997).
\bibitem{KHP99} S. Kwon, W. Hwang, and H. Park, Phys. Rev. E {\bf 59}, 4949 (1999).
\bibitem{KP95} S. Kwon and H. Park, Phys. Rev. E {\bf 52}, 5955 (1995).
\bibitem{CT96} J. L. Cardy and U. C. T\"auber, Phys. Rev. Lett. {\bf 77}, 4780 (1996); J. Stat. Phys. {\bf 90}, 1 (1998).
\bibitem{IT98} N. Inui and A. Yu. Tretyakov, Phys. Rev. Lett. {\bf 80}, 5148 (1998).
\bibitem{DL9} See, e.g., I.D. Lawrie and S. Sarbach, in {\it Phase
Transitions and Critical Phenomena}, edited by C. Domb and J.L. Lebowitz
(Academic Press, London, 1984), Vol. 9.
\bibitem{OM08} G. \'Odor and N. Menyh\'ard, arXiv:0804.3092.
\bibitem{PP07} S.-C. Park and H. Park, Phys. Rev. E {\bf 76}, 051123 (2007).
\bibitem{OMSM98} G. \'Odor, J. F. F. Mendes, M. A. Santos, and M. C. Marques, Phys. Rev. E {\bf 58}, 7020 (1998).
\bibitem{PP08b} S.-C. Park and H. Park (unpublished).
\end{thebibliography}
\end{document}